\documentclass{article}
\title{Simulations of Compton Sequencing with the GRETINA Detector}
\author{Dr. Robert Crabbs \\ University of California, Berkeley \\
	\and 
	Dr. I-Yang Lee \\ Lawrence-Berkeley National Laboratory \\
	\and 
	Dr. Kai Vetter \\	Lawrence-Berkeley National Laboratory \\
	}
\date{\today}

\usepackage{amsmath} 
\usepackage{bm} 
\usepackage{graphicx} 
\usepackage{pdfpages} 
\usepackage{textcomp} 
\usepackage{color} 
\usepackage{multicol,caption} 
\usepackage{geometry} 
\usepackage{hyperref} 

\hypersetup{
  colorlinks   = true, 
  urlcolor     = blue, 
  linkcolor    = black, 
  citecolor   = red 
}
\newcommand{\abs}[1]{\lvert#1\rvert}
\newcommand{\norm}[1]{\lVert#1\rVert}
\newcommand{\scinotation}[2]{#1 $ \times $ 10\textsuperscript{#2}}

\newenvironment{multicolfigure}
  {\par\medskip\noindent\minipage{\linewidth}}
  {\endminipage\par\medskip}

\begin{document}

\maketitle

\begin{abstract}
GRETA, the \textbf{G}amma-\textbf{R}ay \textbf{E}nergy \textbf{T}racking \textbf{A}rray, is an array of highly-segmented HPGe detectors designed to track $ \gamma $-rays emitted in beam-physics experiments. Its high detection efficiency and state-of-the-art position resolution enable it to reject Compton background and also sequence detected interactions via Compton kinematics. In this paper, we use simulated photon tracks to estimate how well interactions can be sequenced in the GRETA detector. This lays the groundwork for subsequent $ \gamma $-ray imaging applications such as nuclear lifetime measurements. 
\end{abstract}

\begin{multicols}{2}


\section{Introduction}
\label{sec:introduction}

Gamma-ray tracking \cite{gamma_ray_tracking_detectors} \cite{new_concepts_in_gamma_detection} is a major advance in gamma-ray spectroscopy. A 4$ \pi $ tracking-array would be a powerful instrument for a broad range of experiments in low-energy nuclear science \cite{nsac_long_range_plan} \cite{nupecc_long_range_plan}, especially for the nuclei far from the line of stability. Developments of these instruments are underway \cite{agata_and_greta} both in the US (GRETINA/GRETA) \cite{gretina_performance_whitepaper} \cite{gretina_spectroscopy_performance} \cite{greta_website} and Europe (AGATA) \cite{agata_whitepaper} \cite{agata_website}. The GRETA collaboration has built a partial array, called the Gamma-Ray Energy Tracking IN-beam Array (GRETINA). The array's primary purpose is for high-efficiency, high-precision $ \gamma $-ray spectroscopy. Energy spectra of the detected photons can help us to better understand the underlying nuclear structure of the reaction products. 
\par
GRETINA's position sensitivity allows experimenters to sequence photon tracks via Compton kinematics. \cite{gamma_ray_tracking_opportunities} This tracking is used, among other things, to estimate the emission angle of a gamma ray relative to its parent nucleus's velocity. This is key for spectroscopy because it gives a way to correct Doppler shifts in lab-frame photon energy. In a beam experiment, an accelerator delivers high-energy projectiles to a thin target to produce excited nuclei through fusion or other nuclear reactions. The product nuclei proceed out of the target in a highly-collimated ``recoil'' beam traveling at relativistic speeds (typically 0.05-0.4c, depending on the beam energy and the kinematics of the nuclear reaction). The recoil nuclei then de-excite a short time later and emit one or more characteristic gamma-rays. Some of these photons will subsequently interact in the detector, leaving a track with one or more discrete energy depositions (called ``hits''). 
\par
Because the parent nuclei are moving relativistically in the lab-frame, the Doppler-shift of the detected $ \gamma $-rays poses a challenge to spectroscopy. This shift can be quite significant (sometimes multiple hundreds of keV), with the lab-frame energy $ E_{Lab} $ given by:
\begin{equation}
\label{eq:energy_calc_doppler_shift}
E_{Lab} ~=~ \frac{E_{CM}}{\gamma (1 - \beta \cos{\phi})}
\end{equation}
where $ E_{CM} $ is a $ \gamma $-ray's known CM-frame energy, $ \beta c $ is the parent nucleus's lab-frame speed along the beam axis, $ \gamma $ is the corresponding Lorentz factor, and $ \phi $ is the emission angle relative to the parent's velocity (as measured in the lab frame). 
\par
GRETINA's position-sensitive modules allow us to correct for these Doppler shifts by providing the locations of photon interactions in the detector volume. We can estimate the emission angle $ \phi $ using the path from the beam target to the \textit{first} hit in the detector. Using this emission angle, we can then transform the measured lab-frame photon energy $ E_{Lab} $ back into its CM-frame energy $ E_{CM} $.
\par
Knowing the interaction sequence is useful for another reason -- reduction of Compton background. Previous-generation gamma detector arrays use Compton suppression layers to reject partial-energy counts. In such setups, detector regions are surrounded by a thick shell of secondary detector material. If a photon scatters out of the sensitive HPGe region and subsequently interacts in the secondary layer, the array rejects the event as Compton background. This is key for accurate and precise measurement of photopeak energies, especially in complex spectra with multiple peaks. (Note that the peaks may overlap by tens of hundreds of keV due to the Doppler-shifting discussed above. This makes it more difficult to distinguish Compton background from Doppler-shifted photopeak counts, hence the need for Compton suppression.) The downside to this approach is a large reduction in counts. The methods described in this paper allow for a more selective discrimination of Compton background.
\par
Furthermore, GRETINA's excellent position- and energy-resolution also enable $ \gamma $-ray imaging. \cite{compton_imaging_in_gretina} \cite{doppler_imaging_in_gretina} If we can accurately reconstruct the photon source distributions from in-beam experiments, we can then measure nuclear lifetimes more efficiently than is possible with current methods such as RDM. \cite{lifetime_measurement_74rb} \cite{rdm_triplex_plunger} \cite{lifetime_measurements_in_gretina} 
\par
Our goal here is to understand performance for sequencing photon interactions detected in GRETINA's segmented crystal array. This is key for using GRETINA in the above-mentioned applications.

\section{Compton Sequencing}
\label{sec:compton_sequencing}

There is a very important caveat -- we do not know the sequence of hits in a photon track. Consider that a 1.0 MeV photon traveling through HPGe has a mean-free-path of several centimeters, and this distance decreases exponentially with photon energy. Moving at the speed of light, then, the photon covers the distance between interactions in mere picoseconds (1.0 cm = 33 ps). Drift times in HPGe are on the order of 100's of nanoseconds, and GRETINA's 100MHz digitizer takes samples every 10 ns. Because of this, the detector cannot distinguish the time sequence of interactions directly -- all the photon hits, from the initial scatter to the final photoabsorption -- appear to happen more or less simultaneously. Figure \ref{fig:sequencing_problem} illustrates the general sequencing problem, which will be discussed in greater detail in Section \ref{sec:compton_sequencing}.
\par
Perhaps the simplest approach is to assume that the first hit is the one with the highest energy deposition. Based on Compton kinematics and scattering cross-sections, this assumption is valid much of the time -- though it is not a certainty. For example, a 1.0 MeV photon that scatters at 30{\textdegree} deposits 208 keV in the detector. The scattered 792 keV photon then has a 36\% chance to make a greater deposit with the \textit{next} hit, resulting in a false identification.

\begin{multicolfigure}
\centering
\includegraphics[width=1.0\textwidth]{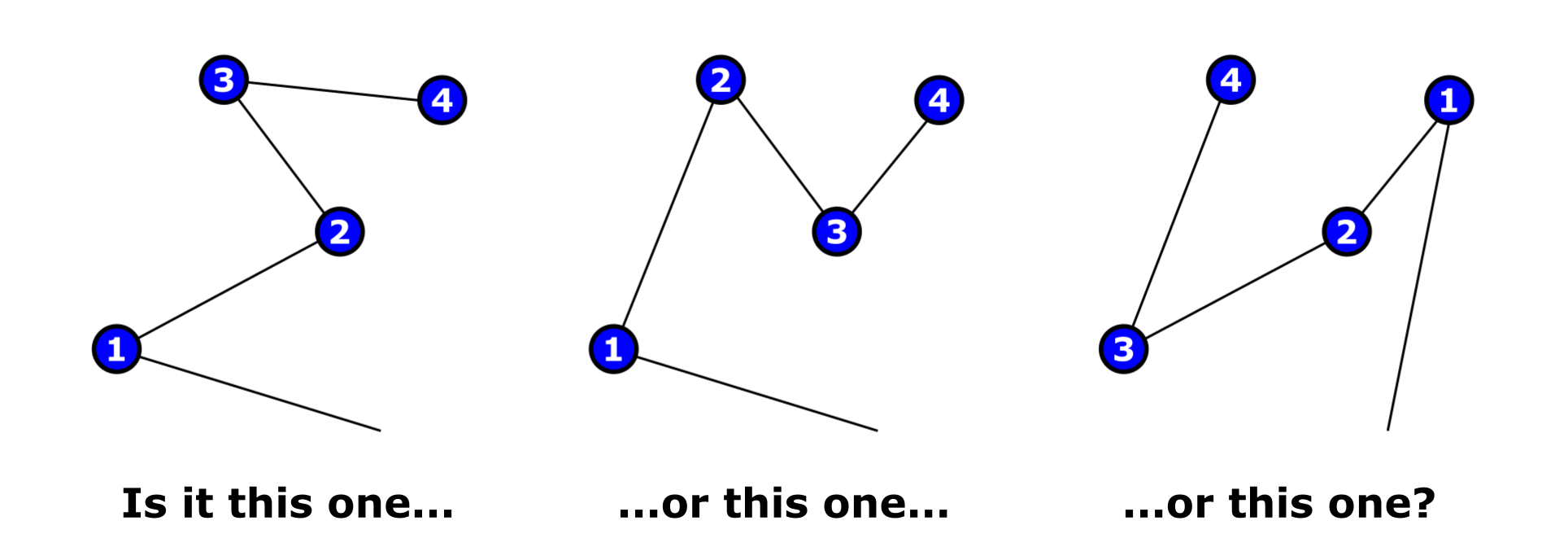}
\captionof{figure}{Illustration of the sequencing problem}
\label{fig:sequencing_problem}
\end{multicolfigure}

\begin{multicolfigure}
\centering
\includegraphics[width=1.0\textwidth]{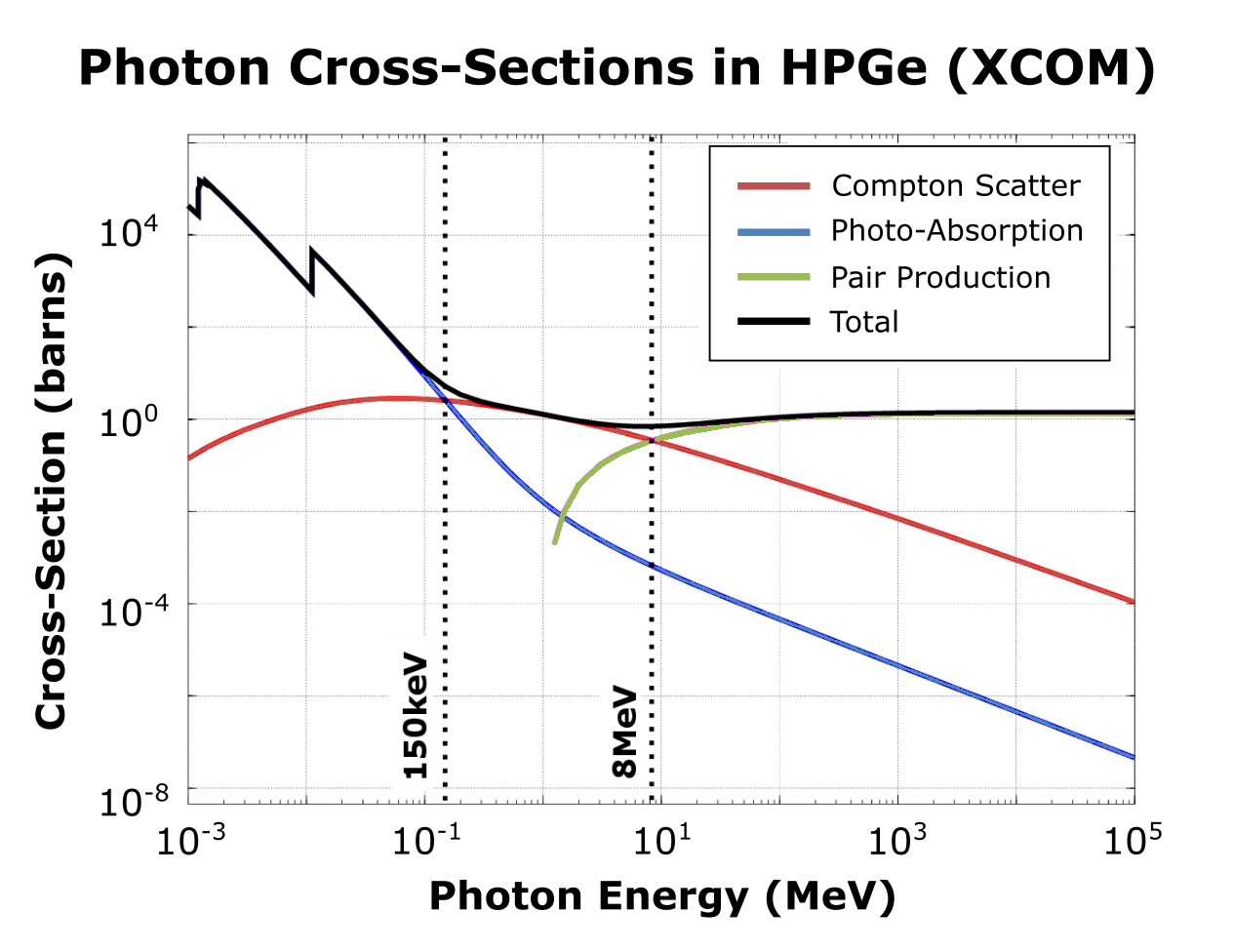}
\captionof{figure}{$ \gamma $-ray cross-sections in HPGe}
\label{fig:gamma_xs_in_hpge}
Compton dominates for 150 keV $ < E_{\gamma} < $ 8 MeV
\\~
\end{multicolfigure}

\par
While we cannot measure the sequence directly, we can deduce it using the energy and position sensitivity of the detector. Recall that photons undergo three main types of interactions in detectors: pair production, photoabsorption, and Compton scattering. The kinematics of this last type of interaction enables a technique called Compton Tracking, which is now widely used in astrophysics, nuclear security, and other scientific work.
\par
Consider a photon track with $ N_{hits} $ total interactions. For each of these interactions, we measure its 3D coordinates $ \bm{X_j} = (x_j, y_j, z_j) $ and energy deposition $ {\Delta}E_j $. $ \bm{X_1} $ and $ \Delta E_1 $ correspond to the first hit in the detector, $ \bm{X_2} $ and $ \Delta E_2 $ with the second, and so on. The total track energy is defined as $ E_{tot} = \sum \Delta E_j $. Below a photon energy of about 4.0 MeV, pair production is rare (refer to Figure \ref{fig:gamma_xs_in_hpge} for a comparison of relative cross-sections in HPGe \cite{xcom_cross_sections}). For energies between around 200 keV and 4.0 MeV, then, we can assume the track is composed of one or more Compton scatters, possibly ending with a photoabsorption. 
\par
The kinematics of Compton scattering are well-defined. That is, if we know the incident photon energy and the scattering angle, we can deduce the outgoing photon energy and the energy imparted to the scattering electron. This energy deposition is measured by the detector and is equal to: 
\begin{align}
{\Delta}E &= E_i - E_f \\
&= E_i - \frac{E_i}{1 + \frac{E_i}{{m_e}c^2}(1 - \cos{\theta})}
\end{align}
Alternately, if we know the incident photon energy and the energy deposition for the scatter, we also know the cosine of the corresponding scattering angle:
\begin{align}
\label{eq:mu_calc}
\mu &= \cos{\theta} \\
&= 1 - {m_e}c ^2 \left( \frac{1}{E_f} - \frac{1}{E_i} \right) \\
&= 1 - \frac{{m_e}c^2{\Delta}E}{E_i (E_i - {\Delta}E)}
\end{align}
Note that there is a one-to-one mapping of energy deposition to scattering angle -- no two angles will yield the same energy deposition for the same initial photon energy. This fact is at the core of Compton sequencing. 
\par
Suppose the first $ N_{cs} = N_{hits} - 1 $ interactions are all Compton scattering. In beam experiments, we don't necessarily know where the photon originated from, so we can't (yet) say anything about the original photon heading vector $ V_0 $, from the emission point to the first hit in the detector. However, the locations of subsequent hits are known. We can define the photon's incidence vector on $ \bm{X_{j}} $ as $ \bm{V_j} = \bm{X_j} - \bm{X_{j-1}} $. If $ \bm{X_j} $ is not the last hit in the track, we can further define the outgoing photon vector at $ \bm{X_{j}} $ as $ \bm{V_{j+1}} = \bm{X_{j+1}} - \bm{X_j} $. Point $ \bm{X_j} $ represents a Compton scatter, and the \textit{measured} scattering angle is equal to:
\begin{equation}
\label{eq:mu_meas}
\mu_{meas,j} = \frac{\bm{V_j} \cdot \bm{V_{j+1}}}{\norm{\bm{V_{j+1}}} \norm{\bm{V_j}}}.
\end{equation}
In other words, $\mu_{meas,j}$ is the cosine of the angle between the incident and outgoing photon directions as given by the coordinates of points $ \bm{X_{j-1}} $ to $ \bm{X_{j+1}} $. This gives us a direct comparison to the theoretical value given by the incident photon energy and energy deposition at $ \bm{V_j} $. Note that because a scattering interaction is anchored by three points, this approach necessarily only applies to tracks with three or more interactions. For a track of length $ N_{hits} $, we only have $ N_{hits} - 2 $ anchored scatterings for which the comparison to theory can be made. So, if the if the source location is unknown, we need at least 3 total hits to sequence a track.
\par
As discussed above, we do not know the track sequence \'{a} priori. But using the combination of theoretical Compton kinematics and the measured coordinates and energy depositions, we can decide which sequences would not make physical sense. Essentially, we can simply try all \textit{possible} sequences and find the one that best matches the Compton formula. This means evaluating a Figure of Merit (FoM) for each permutation of the track sequence.
\par
This approach can become computationally-intensive, because the number of possible sequences scales factorially with the length of the track. An 8-hit track (5,040 permutations) takes roughly 210 times longer to sequence than a 4-hit track (24 permutations). A single track with 12 or more hits can take an hour to process on a mid-range computer from 2015. However, these longer tracks are also fairly rare at the energies where Compton tracking applies (below 4.0 MeV). For example, tracks with 8 or more hits account for less than 1.7\% of all full-energy tracks at 1.0 MeV (Figure \ref{fig:track_breakdown_by_length}). We have chosen here to disregard tracks with 7 or more interactions.

\begin{multicolfigure}
\centering
\includegraphics[width=1.0\textwidth]{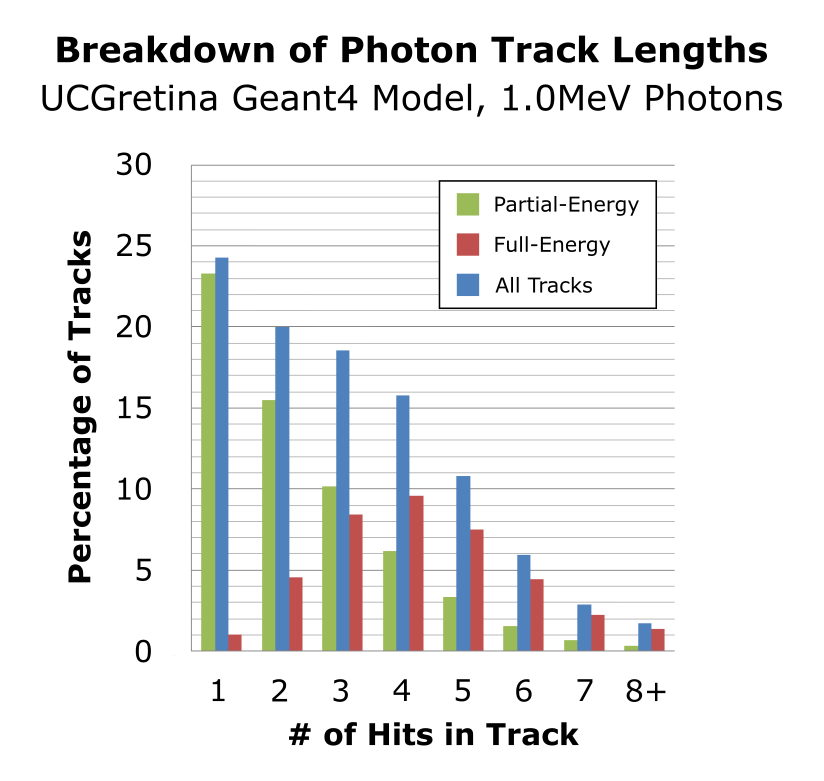}
\captionof{figure}{Track breakdown by length}
(Relative errors are all within 0.15\%)
\label{fig:track_breakdown_by_length}
\end{multicolfigure}

We can also see that 1.0 MeV photons typically scatter multiple times before being of low enough energy for photoabsorption. A large majority of the 1- and 2-hit tracks correspond to partial-energy detections (shown in green). Full-energy detections (shown in red) become more common above three hits. Because we cannot sequence partial-energy or 1-2 hit tracks with Compton kinematics, and because we disregard tracks with 7 or more hits for computational efficiency, \textit{sequenceable} tracks account for only about 30\% of the total at 1.0 MeV.
\par
There are multiple figures of merit we can use to deduce the sequence of interactions. One common FoM compares the cosines of the measured and theoretical scattering angles (Eq. \ref{eq:mu_meas} and Eq. \ref{eq:mu_calc}, respectively) for each scattering in the track:
\begin{equation}
FoM = \frac{1}{ (N_{sc} - 1)}{\sum_{j=2}^{N_{hits}-1} (\mu_{meas,j} - \mu_{calc,j})^2 }
\end{equation}
This figure of merit is unitless and normalized to the number of scattering points being compared. $ \mu_{calc,j} $ is the angle predicted by the incident photon energy and measured energy deposition (Eq. \ref{eq:mu_meas}), and depends on $ E_{i,j-1} $, the incident photon energy at $ \bm{X_j} $. $ E_{i,j-1} $ is calculated from the total energy of the track, minus all previous energy depositions:
\begin{align}
E_{i,j-1} &= E_{tot} - \sum_{k=1}^{k=j-1} {\Delta}E_k \\
&= \sum_{k=j}^{k=N_{hits}} {\Delta}E_k
\end{align}
Incorrect sequences generally produce poor fits to the above equation. Suppose we are testing a 4-hit track sequence with hits \#2 and \#3 swapped. (Figure \ref{fig:compton_sequencing}) In this case, the incident headings on $ \bm{X_2} $ and $ \bm{X_3} $ are both incorrect. This means $ E_2 = {\Delta}E_2 + {\Delta}E_4 $ instead of the correct $ E_2 = {\Delta}E_3 + {\Delta}E_4 $, and the calculated scattering angles at $ \bm{X_2} $ and $ \bm{X_3} $ will reflect the inaccuracy. 

\begin{multicolfigure}
\centering
\includegraphics[width=1.0\textwidth]{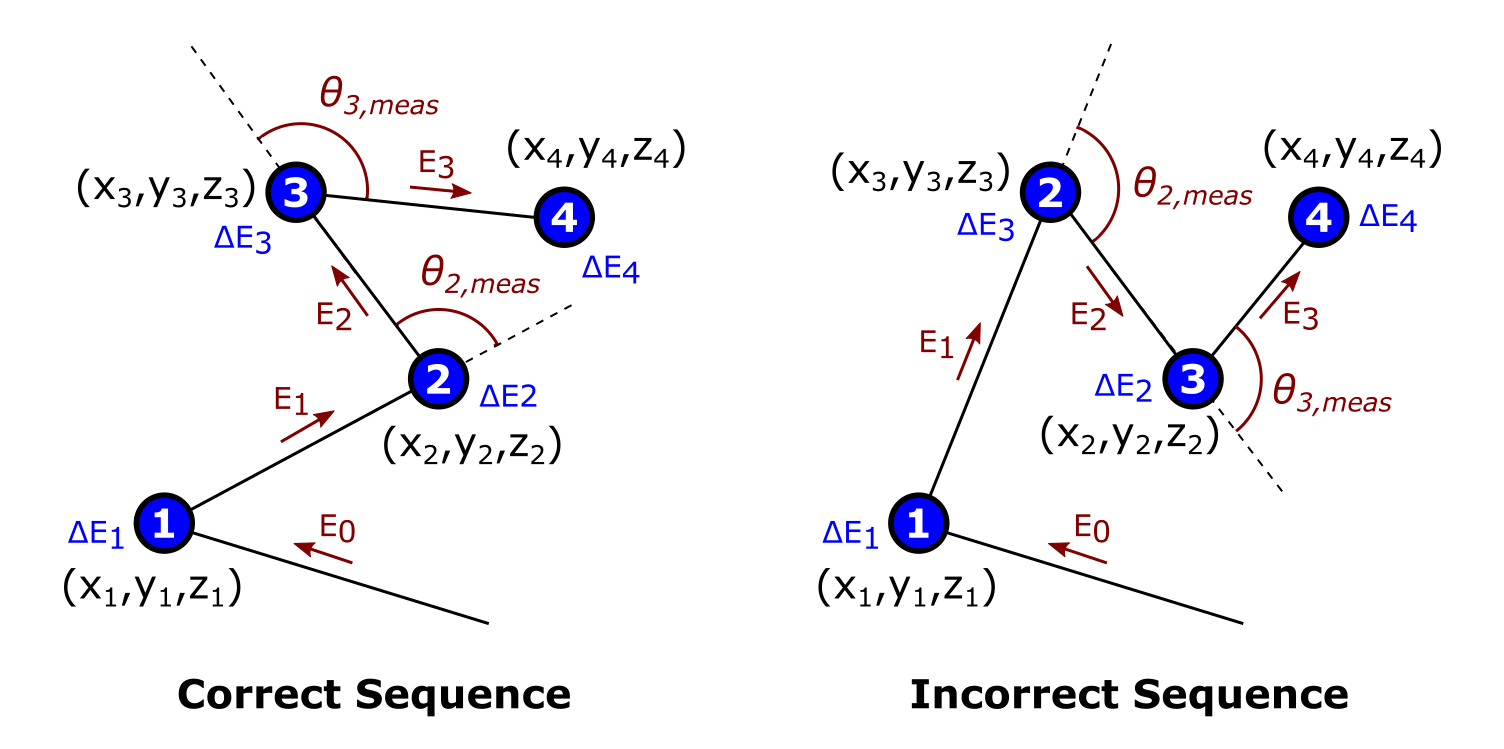}
\captionof{figure}{Evaluating two different sequences}
\label{fig:compton_sequencing}
\end{multicolfigure}

\par
We can define an alternative FoM using the energy depositions instead of scattering angle:
\begin{equation}
\label{eq:sequencing_FoM_angle}
FoM = \frac{1}{ (N_{cs} - 1)}{\sum_{j=2}^{N_{hits}-1} \frac{({\Delta}E_{meas,j} - {\Delta}E_{calc,j})^2}{{\Delta}E_{meas,j}^2} }
\end{equation}
Here, $ {\Delta}E_{meas,j} $ is the measured energy deposition at $ \bm{X_j} $. $ {\Delta}E_{calc,j} $ is the measured energy deposition at $ \bm{X_j} $. To obtain a unitless figure of merit, the observed difference in measured vs. calculated energy depositions is divided by $ E_{meas,j} $. Normalizing in this fashion weights each scatter equally, regardless of the magnitude of the energy deposition. However, in most cases we want to prioritize accuracy on the first (and sometimes second) interactions. As noted earlier, these early interactions tend to have much higher energy depositions than subsequent hits in the track. Normalizing by total track energy weights these high-energy scatters over the small-energy ones near the end of a track:
\begin{equation}
\label{eq:sequencing_FoM_energy}
FoM = \frac{1}{ E_{total}^2 (N_{cs} - 1)}{\sum_{j=2}^{N_{hits}-1} ({\Delta}E_{meas,j} - {\Delta}E_{calc,j})^2 }
\end{equation}
This is the normalization and figure of merit used to perform sequencing in this study. However, there may be better choices that would provide better discrimination between Compton background and photopeaks counts. One could test different FoMs and evaluate them based on a desirable metric, such as the reliability of sequencing the first 2 hits in a track (discussed further in Section \ref{sec:sequencing_effectiveness}).

\section{Interaction Grouping}
\label{sec:interaction_grouping}

Before moving on, it is important to discuss the non-trivial challenge of identifying which hits belong to which photon track. During an in-beam experiment, we may register millions of separate photon interactions over the course of an hour. GRETINA will return a separate timestamp, energy deposition, and set of 3D coordinates for each one.
\par
Recall that each crystal in the GRETINA array is heavily segmented into 36 individual regions. These segments are relatively small (approx. 1.5 cm in depth), which means it is unlikely for a high-energy gamma-ray to deposit its full energy into a single crystal segment. When a photon scatters from one segment into another, GRETINA records separate ``events'' for each triggered segment. Note that an \textit{event} can contain multiple \textit{hits} -- this occurs when the photon interacts multiple times within a single crystal segment. To perform sequencing, we need to group interactions from nearby segments to re-form a photon track as a whole. We refer to this process as interaction ``grouping'' or ``clustering''.
\par
\begin{figure}
\centering
\includegraphics[width=0.8\textwidth]{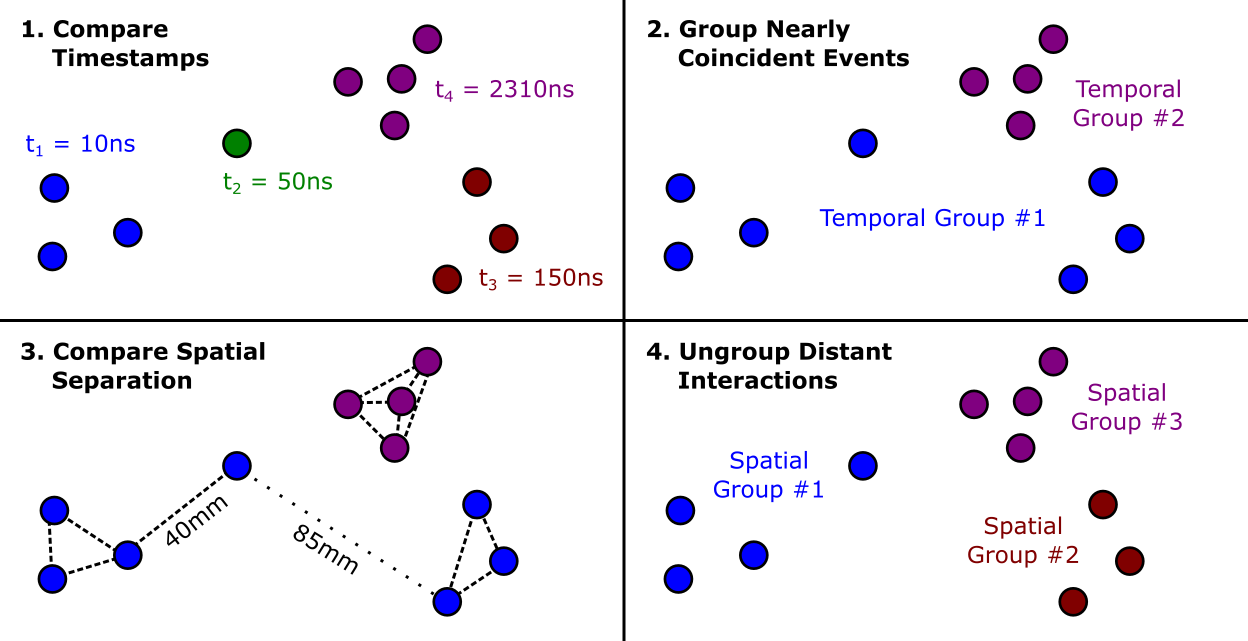}
\caption{Illustration of interaction grouping}
\label{fig:interaction_grouping}
\end{figure}

The clustering process consists of several steps, illustrated in Figure \ref{fig:interaction_grouping}. First, interactions are grouped by trigger time. GRETINA's electronics sample at 100 MHz, with all segments synced to the same clock. Because the clocks are synced, we can then apply a timing gate to group "simultaneous" events and hits. (Determining multiple hits within a segment is very challenging work. This is currently one of the primary focuses of the signal decomposition team. \cite{signal_decomp_in_GRETINA} \cite{challenges_in_signal_decomp})
\par
A secondary effect arises from time gating. Coincident gamma rays will be falsely clumped together, producing a single track with an unphysical energy. To mitigate problems from such event pileup, a secondary layer of clustering can be applied to re-separate interactions by spatial proximity. If two interactions are more than several MFP's apart, then they are unlikely to belong to the same track. The software algorithm to accomplish the separation is somewhat complicated, but the end result is that temporally-clustered events are regrouped into a larger number of spatially-distinct tracks.
\par
Implementations differ based on how ``proximity'' is defined. We can define it in a solid-angle sense, or by using the 3D lab coordinates of each photon hit. Our analysis uses this second approach. We set a hard threshold on the physical distance between points in a given group. As long as a point is close enough to \textit{any} other point in the group, it remains in the group. The approach is somewhat simplified -- it is based on absolute distance between points. This can incorrectly ungroup tracks for which the photon scattered from one detector into another by crossing GRETINA's inner cavity. A more correct algorithm would use only the travel distance in the HPGe itself. 
\par
The choice of timing gate and spatial clustering threshold can be optimized through energy spectra. We would expect interaction grouping to improve P/T ratios, because it assembles photopeak tracks from events that would otherwise have fallen into the Compton background. Spatial re-separation recovers some of the photopeak counts that are erroneously grouped with other tracks. 
\par
Because the distance between hits depends on cross-section, spatial clustering should vary by photon energy. We used several different calibration sources to study this dependence: \textsuperscript{152}Eu, \textsuperscript{60}Co, \textsuperscript{88}Y, \textsuperscript{22}Na. Table \ref{table:calibration_sources} lists characteristic $ \gamma $-rays and the experimental dataset for each source.
\par
Tables \ref{table:pt_ratios_vs_spatial_grouping_threshold} and \ref{table:pt_ratios_vs_temporal_grouping_threshold} list the P/T ratios obtained for a range of temporal and spatial clustering thresholds. For the spatial threshold study, we kept the temporal threshold constant at 250 ns, while for the temporal threshold study, we kept the spatial threshold constant at 70.0 mm. Peak counts were taken from 10-12 keV regions centered about the known photopeak energies. Background was defined by averaging counts over the regions immediately to the left and right of the peaks, and was then subtracted from the peak counts using a trapezoidal approximation.
\par
Figure \ref{fig:energy_spectra_grouping} illustrates how the peak-to-total (P/T) ratio improves with both types of clustering. The improvements are modest but welcome -- in this case, a 26\% total improvement in P/T and 22\% reduction in the Compton background for a \textsuperscript{60}Co source. The effect of spatial clustering threshold on the P/T ratio is not as pronounced as we had expected, and generally, 300 ns / 80 mm thresholds will work well for all photons between 120 keV and 1.8 MeV. The bolded elements in the tables show where the P/T ratio reaches 99\% of its maximum value for a given peak energy. 

\begin{multicolfigure}
\centering
\includegraphics[width=1.0\textwidth]{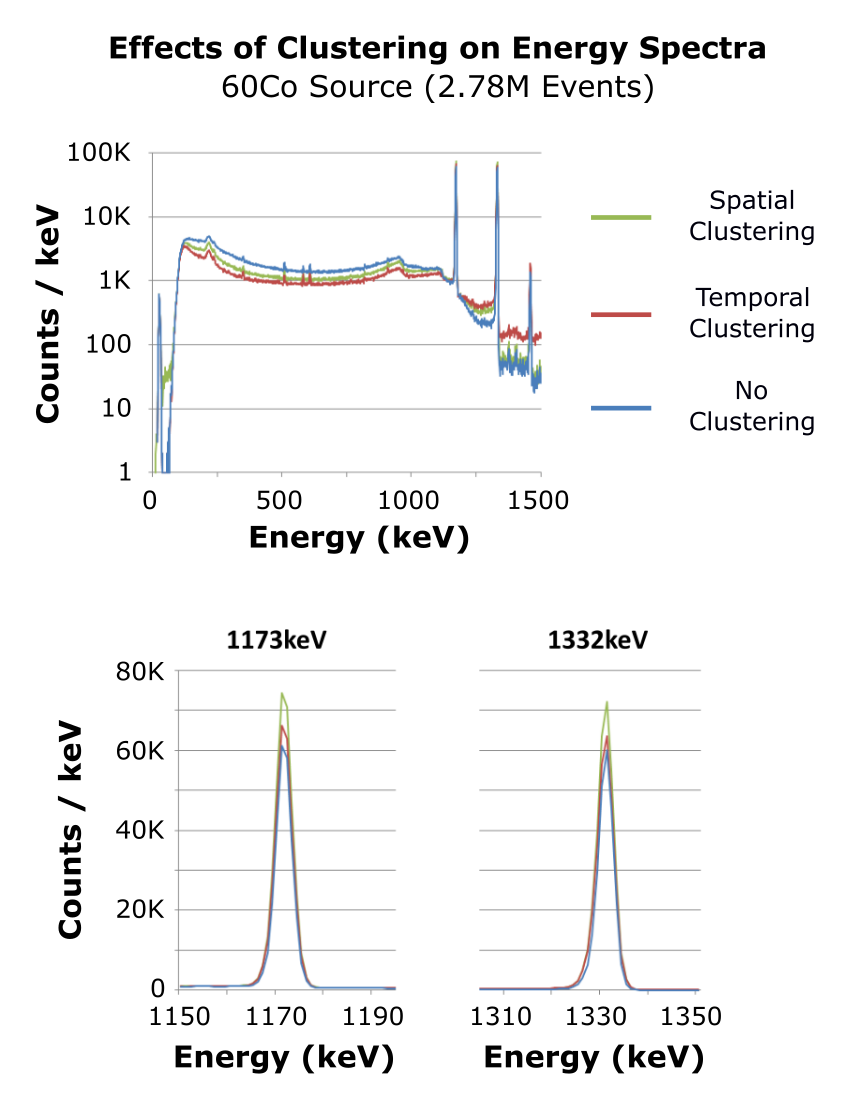}
\captionof{figure}{\textsuperscript{60}Co energy spectra using various}
clustering levels
\label{fig:energy_spectra_grouping}
\end{multicolfigure}

\par
We also looked at signal-to-noise (SNR) ratios, defined as:
\begin{equation}
SNR = \frac{\text{[Gross Peak Counts] - [Background Counts]}}{\text{[Background Counts]}}
\end{equation}
Background under each peak was estimated with a trapezoidal approximation, using average counts to either side of the peak. As we saw, the effects were fairly small for thresholds above 80mm. For example, the SNR for the 1836.1 keV \textsuperscript{88}Y peak showed a maximum of 68.8 at 40 mm, but fell to 57.1 at 120 mm. Therefore, we conclude that the data quality is not harmed by excessively large clustering thresholds.

\end{multicols}
\begin{multicols}{2}

\section{Sequencing Efficiency and Effectiveness}
\label{sec:sequencing_effectiveness}

Compton sequencing is relatively straightforward on paper, but real-world experimentation introduces some complications. For example, the ability to use Compton kinematics for sequencing depends on photon energy. Higher energies give longer tracks and hence more data to work with, but also more potential sequences from which to choose. Below a certain energy, Compton sequencing cannot be used because too many tracks will have one or two hits. 
\par
Real detectors also have finite position and energy resolution. \cite{position_sensitivity_in_hpge} \cite{greta_detector_performance} This means that the measured interaction coordinates $ \bm{X_j} $ are only approximations of the true locations. Therefore, the measured scattering angles we compute from these positions are themselves uncertain, and the FoMs calculated in Equation \ref{eq:sequencing_FoM_energy} will be nonzero -- even for the \textit{correct} sequence of hits.
\par
Even with perfect detector resolution, though, basic Compton kinematics are also only an approximation. There is an assumption that the photon scatters off of a free electron at rest. However, electrons in real materials have characteristic binding energies that give them nonzero momentum prior to the interaction. This extra momentum may cause a scattered photon to have significantly different energy from that predicted by the Compton formula. (The effect is known as ``Doppler Broadening''. \cite{doppler_broadening}) Because bound electron momentum is inherently stochastic in nature, a deterministic calculation is impossible -- we can never be completely certain of any sequence deduced from kinematics. 
\par
The severity of these problems depends on the application. Mis-sequencing errors in $ \gamma $-ray imaging are discussed in more detail in References \cite{compton_imaging_in_gretina} and \cite{doppler_imaging_in_gretina}. Doppler-shift corrections are also affected -- a mis-sequenced track changes the calculated emission angle. Consider -- hits in a 1.0 MeV track will typically be a few cm apart. Therefore, a mis-sequenced interaction will result in an angular deviation of 10\textdegree{} or less with GRETINA's 18 cm inner radius. The Doppler correction (from Equation \ref{eq:energy_calc_doppler_shift}) for a 90\textdegree{} emission from a nucleus moving at 0.3c is $ E_{CM}/E_{Lab} = 0.962 $. The same correction for a 80\textdegree{} emission is $ E_{CM}/E_{Lab} = 1.01 $, i.e. a difference of 4.8\%. In this particular case, the corrected photon energy is nearly 50 keV off.
\par
To better quantify the effectiveness of sequencing, we used a detailed Geant4 model to generate simulated photon interactions within the GRETINA array. \cite{UCGretina_model} This allowed us to incorporate realistic detector geometry, detector resolution, materials, and Doppler broadening into our study of Compton sequencing. We used a stationary 1.0 MeV point source located at the center of GRETINA's central cavity, i.e. the origin in the Geant4 model (the $ z $-axis is oriented along the beam line) Each simulated track included the true sequence, locations, and energy depositions for each associated photon interaction. Note that the simulated output produced \textit{pre-grouped} photon tracks -- we did not need to worry about clustering here. To reflect the response of a real detector, we randomly re-ordered the interactions in each simulated track, then added uniform Gaussian noise to the position and energy of each hit. The intentionally randomized tracks were then run through the sequencing algorithm. Lastly, we compared the known tracks with the Compton-sequenced versions to determine the frequency of sequencing errors.
\par
Correctly identifying the first interaction is key for Doppler-shift corrections and Doppler-shift imaging. Additionally, the second interaction is important when using the sequence for Compton imaging. We therefore looked at error probabilities for mis-sequencing the first hit alone, the first and second hits together, and for the track as a whole. Only \textit{full-energy} tracks were accepted for processing (\textpm{}6 keV), or approximately 39\% of the 100K simulated tracks. (Refer back to Figure \ref{fig:track_breakdown_by_length} for the breakdown of the simulated tracks by length and energy.) 2-hit tracks were also included, and were sequenced by assuming the higher-energy hit came first.
\par
Tables \ref{table:sequencing_quality_vs_position_resolution} and \ref{table:sequencing_quality_vs_energy_resolution} show mis-sequencing probabilities as functions of simulated detector resolution for tracks with 2 to 6 hits. Table \ref{table:sequencing_quality_vs_track_length} gives similar numbers for a typical detector as a function of the photon track length. For these runs, detector resolution was kept constant with standard deviations of $ \sigma_{xyz} = $ 3.0 mm and $ \sigma_E = $ 2.0 keV for position and energy resolution, respectively. With perfect simulated detector resolution (i.e. no uncertainty about the locations or energy of each interaction), we were able to correctly reconstruct virtually all full-energy interaction sequences correctly. 
\par
As discussed, Compton sequencing requires three or more photon hits to work. The more hits in a photon track, the more scattering points we'll have to compare to kinematics. Intuitively, this should allow us to sequence longer tracks with greater certainty. However, longer tracks also have more potential ways to be mis-sequenced. (All things being equal, the probability of randomly guessing a 6-hit sequence correctly is only 0.14\%, versus 17\% for a 3-hit track.) Longer tracks also have more low-energy hits, which are disproportionately affected by the detector's finite position and energy resolution. While we might be able to sequence the first few interactions reliably, the tail end of a photon track is more difficult. 
\par
In addition, the percentage of sequenceable tracks depends on photon energy. Low-energy photons tend to leave shorter tracks with fewer scatters. On the other hand, high-energy photons have a lower probability of full-energy deposition in the detector. Because we assume full-energy deposition in sequencing, Compton background counts may not sequence properly.
\par
The results of Table \ref{table:sequencing_quality_vs_track_length} suggest some important points. First, given 3.0 mm position- and 2.0 keV energy-resolution in the detector, nearly half of all full-energy tracks -- 46\% -- show an error in the first two hits. This has serious repercussions for Compton imaging -- without some kind of filter, nearly half of all tracks will contribute noise to the final image. The situation is somewhat improved for Doppler-shift correction and imaging (Reference \cite{doppler_imaging_in_gretina}), both of which only require the first hit in a track. Still, a full 1/3rd of tracks yield incorrect results here. 
\par
Track length also has a very significant impact. For example, the ratio of correctly-sequenced to mis-sequenced 2-hit tracks here is 5.5, whereas for 4-hit tracks it is only 0.78. Because of this, it may be advantageous not to use Compton sequencing in certain situations (e.x. for low-energy photon emissions). If we only need to get the first interaction right, it may be more accurate to simply assume it is the highest-energy hit.
\par
Not having that first scattering point available for FoM calculations is a significant limitation to our approach. As discussed earlier, the first hit usually corresponds to the highest energy deposition. This also means that the first hit is also likely to contribute the largest amount to the sequencing FoM. At high photon energies, small deviations in scattering angle can yield 10's of keV difference in energy deposition. Not so with lower-energy photons. Since our FoM is normalized to the \textit{total} track energy, this means sequencing mistakes for low-energy hits are less noticeable in the final FoM. This is by design, since we want to prioritize getting the first (i.e. higher-energy) hits correct.
\par
Table \ref{table:sample_sequenced_track} shows a sample mis-sequenced track. Note that Compton sequencing on this 6-hit track gave an incorrect location for the first interaction, despite \textit{perfect} detector position and energy resolution. This rare occurrence would have been avoided had we been able to use our \'{a} priori knowledge of the source location (the center of GRETINA's central cavity). The FoM contribution from the first hit in the computed sequence is quite high relative to the other FoM contributions. The remaining hits happened to be arranged in such a way that the subsequent hits still matched expectations from Compton kinematics fairly well. This may sound wildly improbable at first glance. However, it can actually take fairly large deviations in angle to skew the energy deposition more than a few keV. Consider that the measured scattering angle is calculated:
\begin{align}
\theta_{meas,i} &= \cos^{-1}{(\bm{\hat{V}_{i-1}} \cdot \bm{\hat{V}_{i}})} \\
\bm{\hat{V_i}} &= (\bm{X_{i+1}} - \bm{X_i}) / \abs{\bm{X_{i+1}} - \bm{X_i}}
\end{align}
Again, $ \bm{\hat{X_0}} $ is the photon emission point and $ \bm{\hat{X_1}} $ is the first hit in the detector. We find $ \theta_{meas,1} $ = 2.285 radians at $ \bm{\hat{X_1}} $ in the true sequence, while the angle calculated by Compton kinematics is $ \theta_{calc,1} $ = 2.204 radians. This is a difference of 3.51\%, yet the corresponding difference in measured vs. calculated energy deposition is only -0.93\% (757.012 keV vs 764.042 keV). Even a 10.00\% difference in the initial scattering angle would only produce a 3.00\% in energy deposition (779.726 keV vs. 764.042 keV).
\par
Lastly, we explored several different sequencing Figures-of-Merit:
\begin{align}
FoM_1 &= \sum (\mu_{meas,i} - \mu_{calc,i})^2 / N_{cs} \\
FoM_2 &= \sum (\Delta E_{meas,i} - \Delta E_{calc,i})^2 / (N_{cs} E_{total}^2) \\
FoM_3 &= \sum (\Delta E_{meas,i} - \Delta E_{calc,i})^2 / (N_{cs} \Delta E_i^2) \\
FoM_4 &= \sum (\Delta E_{meas,i} - \Delta E_{calc,i})^2 / (N_{cs} E_{total}^2)
\end{align}
FoM 4 is somewhat different than the others -- we take an educated \textit{guess} at the photon origin $ \bm{X_0} $, and add it to the theoretical sequence for evaluation. In lifetime experiments, for example, we might assume photons are emitted at the origin even though the source is not stationary. This guess might be off by several centimeters, but this may result in only a small angular error (and smaller error in deposited energy). This approach therefore provides an additional data point to compare to Compton kinematics. FoMs 1, 2, and 3 apply to hits 2 through $ N_{hits} - 1 $, while FoM 4 applies all hits 1 to $ N_{hits} - 1 $. 
\par
Table \ref{table:sequencing_quality_vs_FoM_definition} summarizes the results. We found that comparing energy depositions and normalizing to the total track energy (FoM 2) provides better results than comparing cosines of scattering angles (FoM 1). Normalizing to \textit{total} track energy is also better than normalizing to the measured energy deposition at each hit (FoM 3). 
\par
Note that FoM 4 did offer some improvement for (3.0 mm, 2.0 keV) detector resolution. For this initial test, we placed our guess for $ \bm{X_0} $ at (0.0 mm, 0.0 mm, 60.0 mm), or 60.0 mm away from the true source location. In lifetime measurements the emission points can be spread over a wide range (100 mm or more), so this seemed a reasonable first attempt. While we did not study the FoM4 approach in great detail, the performance numbers in Table \ref{table:sequencing_quality_vs_FoM_definition} are promising. The results improved markedly as the guess for $ \bm{X_0} $ came closer to the true source location at (0.0 mm, 0.0 mm, 0.0 mm), approaching near-100\% accuracy when the guess was spot-on. However, this FoM definition can actually \textit{degrade} sequencing performance if the guess is too far off. We therefore chose to use FoM 3 as the working definition for subsequent imaging work.
\par
An additional observation about sequencing -- a good Figure-of-Merit \textit{ought} to be correlated with whether or not a track deposits its full energy in the detector. One would expect partial energy tracks to match Compton kinematics relatively poorly and thus have a worse FoM than a full-energy track. Because of this, the Figure-of-Merit is a potential way to reduce background counts, thereby improving P/T ratios for imaging or spectroscopy purposes. Receiver-Operating Characteristic (ROC) curves are a standard method for evaluating the classification effectiveness of a given filter. \cite{ROC_curves} We are interested in the system sensitivity (true positive rate) and specificity (true negative rate) as functions of a maximum FoM limit. 
\par
To explore this, we ran a series of tests using increasingly-stringent sequencing FoM thresholds. Tracks were classified as ``true positives'' if they were full-energy events and the first two hits were sequenced correctly. ``True negatives'' were any partial-energy or incorrectly-sequenced track. Unfortunately, as Figure \ref{fig:sequencing_FoM_ROC_curves} shows, the method's sensitivity is not very high for realistic values of detector position resolution. The ROC curve for (3.0 mm, 2.0 keV) resolution is only slightly above the ``random guess'' line. This is consistent with Table \ref{table:sequencing_quality_vs_photon_energy}, which shows that roughly 45\% of \textit{full-energy} tracks contain sequencing errors in the first two hits. Tightening the sequencing FoM limit greatly reduces the number of accepted counts, but has little affect on the sensitivity. A different choice of FoM might yield more promising results.

\begin{multicolfigure}
\centering
\includegraphics[width=1.0\textwidth]{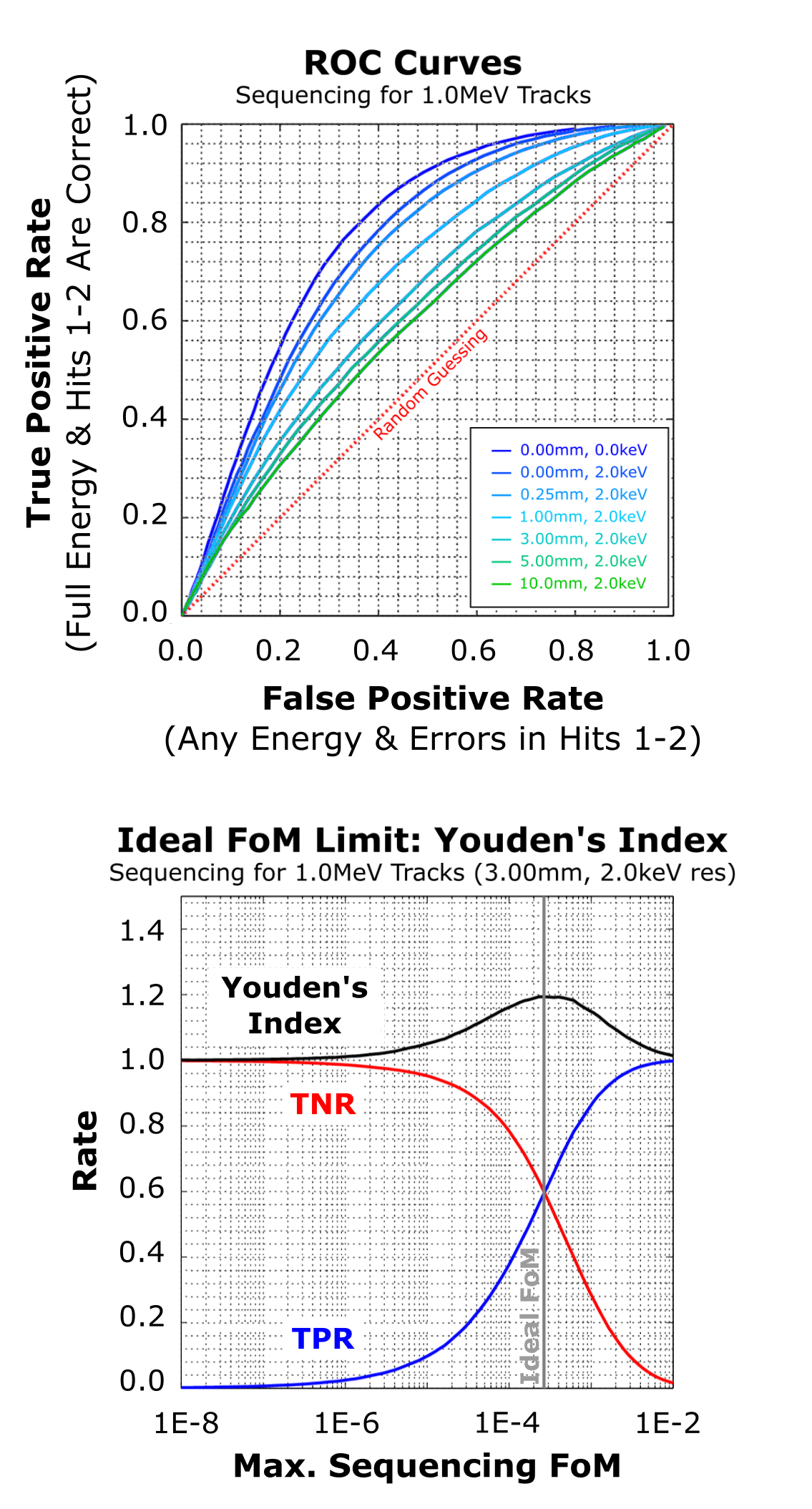}
\captionof{figure}{ROC curves and choosing a}
sequencing FoM limit \\
(TNR = 1 - FPR)
\label{fig:sequencing_FoM_ROC_curves}
\end{multicolfigure}

\par
There are a number of metrics we could use to select an optimal filtering parameter. The Youden index is one such metric, and is given by the sum of the classification sensitivity and specificity. \cite{youdens_index} The maximum value of this index provides a reasonable choice for the parameter value. As we can see in Figure \ref{fig:sequencing_FoM_ROC_curves}, this gives an ``ideal'' FoM limit of \scinotation{2.66}{-4} for 1.0 MeV tracks with a detector of 3.0 mm position- and 2.0 keV energy-resolution.

\section{Future Work}
\label{sec:future_work}

There are multiple directions where we could expand on the research presented here. For example, we limited our study to a single photon energy. 1.0 MeV is a typical emission energy, but sequencing performance is energy-dependent. How strong that dependence is remains unknown.  
\par
The study of sequencing Figures-of-Merits might also be approached more exhaustively. We explored 4 different FoM definitions, the 4th of which introduced a new parameter -- a guess for the position of the emission point. This FoM showed promising performance but also had the potential to degrade performance. What strategies might we employ around this technique for beam experiments, where the source distribution is not a simple point-source?
\par
Lastly, real-world experiments may involve multiple emission lines from the same parent nuclei. We may not have the luxury of a clean, mono-energetic energy spectrum. How sequencing may be impacted by complex, realistic energy spectra is an open question.



\end{multicols}



~\\
~\\

\begin{table}[ht]
\centering 
\begin{tabular}{c c c c}
\textbf{Source} & \textbf{Photon Energies} & \textbf{GRETINA Dataset} \\ [0.5ex]
\hline
\textsuperscript{152}Eu & 121.8, 344.3, 1408.0 keV & MSU\_May2013\_90deg\_\_R29 \\
\textsuperscript{60}Co & 1173.3 keV & MSU\_May2013\_90deg\_\_R16 \\
\textsuperscript{88}Y & 898.0, 1836.1 keV & MSU\_May2013\_90deg\_\_R44 \\
\textsuperscript{22}Na & 511 keV & MSU\_May2013\_90deg\_\_R48
\end{tabular}
\caption{Sources used to optimize clustering thresholds}
\label{table:calibration_sources}
\end{table}

~\\
~\\
~\\
~\\

\begin{table}[ht]
\centering 
\begin{tabular}{c | c c c | c c | c c | c}
\textbf{Hit} & \textbf{X} & \textbf{Y} & \textbf{Z} & \textbf{E\textsubscript{i}} & \textbf{$ \mu_{meas} $} & \textbf{$ \Delta $E\textsubscript{meas}} & \textbf{$ \Delta $E\textsubscript{calc}} & \textbf{FoM} \\
\hline
\multicolumn{9}{c}{\textbf{True Sequence}} \\
\textbf{(Src)} & 0.000 & 0.000 & 0.000 & -- & -- & -- & -- & -- \\
1 & -77.921 & 170.052 & 103.876 & 1000.000 & -0.654642 & 757.012 & 764.042 & \textbf{(9.885E-06)} \\
2 & -77.193 & 168.620 & 96.225 & 242.988 & 0.855815 & 34.633 & 15.591 & 7.252E-05 \\
3 & -81.571 & 165.890 & 87.314 & 208.355 & 0.727474 & 21.966 & 20.837 & 2.549E-07 \\
4 & -81.686 & 166.218 & 86.609 & 186.389 & -0.637781 & 82.652 & 69.705 & 3.352E-05 \\
5 & -81.948 & 166.648 & 87.717 & 103.737 & 0.651133 & 8.124 & 6.861 & 3.189E-07 \\
6 & -81.976 & 166.710 & 87.737 & 95.613 & -- & 95.615 & -- & -- \\ 
\hline
\multicolumn{9}{c}{\textbf{Computed Sequence}} \\
\textbf{(Src)} & 0.000 & 0.000 & 0.000 & -- & -- & -- & -- & -- \\
1 & -77.193 & 168.620 & 96.225 & 1000.000 & 0.632993 & 34.633 & 418.000 & \textbf{(2.939E-02)} \\
2 & -77.921 & 170.052 & 103.876 & 965.367 & -0.944982 & 757.012 & 758.845 & 6.722E-07 \\
3 & -81.948 & 166.648 & 87.717 & 208.355 & 0.885275 & 8.124 & 9.311 & 2.818E-07 \\
4 & -81.686 & 166.218 & 86.609 & 200.231 & -0.637860 & 82.652 & 78.272 & 3.835E-06 \\
5 & -81.571 & 165.890 & 87.314 & 117.579 & -0.021635 & 21.966 & 22.380 & 3.422E-08 \\
6 & -81.976 & 166.710 & 87.737 & 95.613 & -- & 95.615 & -- & -- \\
\hline
\end{tabular}
\caption{Sample Compton-Sequenced Photon Track (1000.0 keV, 6 hits)}
All positions in mm and energies in keV. 
\label{table:sample_sequenced_track}
\end{table}

\begin{table}[ht]
\centering 
\begin{tabular}{c c c c}
\hline \hline
\textbf{Threshold} & \textbf{121.8.1 keV} & \textbf{344.3 keV} & \textbf{511.0 keV} \\ [0.5ex]
\hline
10.0 mm & 0.03568 \textpm{} 0.00013 & 0.01408 \textpm{} 0.00008 & 0.01822 \textpm{} 0.00005 \\
20.0 mm & 0.06614 \textpm{} 0.00017 & 0.03919 \textpm{} 0.00012 & 0.06001 \textpm{} 0.00009 \\
30.0 mm & 0.09503 \textpm{} 0.00021 & 0.06328 \textpm{} 0.00017 & 0.10762 \textpm{} 0.00014 \\
40.0 mm & 0.10613 \textpm{} 0.00023 & 0.07390 \textpm{} 0.00018 & 0.13327 \textpm{} 0.00016 \\
50.0 mm & 0.10999 \textpm{} 0.00024 & 0.07865 \textpm{} 0.00019 & 0.14599 \textpm{} 0.00017 \\
60.0 mm & 0.11215 \textpm{} 0.00024 & 0.08124 \textpm{} 0.00020 & 0.15319 \textpm{} 0.00018 \\
70.0 mm & 0.11332 \textpm{} 0.00024 & 0.08255 \textpm{} 0.00020 & 0.15694 \textpm{} 0.00018 \\
80.0 mm & 0.11388 \textpm{} 0.00024 & 0.08312 \textpm{} 0.00020 & 0.15866 \textpm{} 0.00018 \\
90.0 mm & 0.11411 \textpm{} 0.00025 & 0.08337 \textpm{} 0.00020 & 0.15940 \textpm{} 0.00019 \\
100.0 mm & 0.11414 \textpm{} 0.00025 & 0.08347 \textpm{} 0.00020 & 0.15967 \textpm{} 0.00019 \\
110.0 mm & 0.11410 \textpm{} 0.00025 & 0.08349 \textpm{} 0.00020 & 0.15972 \textpm{} 0.00019 \\
120.0 mm & 0.11403 \textpm{} 0.00025 & 0.08344 \textpm{} 0.00020 & 0.15963 \textpm{} 0.00019 \\ [0.5ex]
\hline \hline
\textbf{Threshold} & \textbf{898.0 keV} & \textbf{1173.3 keV} & \textbf{1332.5 keV} \\ [0.5ex]
\hline
10.0 mm & 0.00888 \textpm{} 0.00003 & 0.00858 \textpm{} 0.00004 & 0.00768 \textpm{} 0.00004 \\
20.0 mm & 0.03443 \textpm{} 0.00007 & 0.03518 \textpm{} 0.00010 & 0.03179 \textpm{} 0.00009 \\
30.0 mm & 0.06885 \textpm{} 0.00011 & 0.07328 \textpm{} 0.00017 & 0.06758 \textpm{} 0.00015 \\
40.0 mm & 0.09160 \textpm{} 0.00013 & 0.09966 \textpm{} 0.00021 & 0.09265 \textpm{} 0.00019 \\
50.0 mm & 0.10461 \textpm{} 0.00014 & 0.11545 \textpm{} 0.00023 & 0.10777 \textpm{} 0.00022 \\
60.0 mm & 0.11236 \textpm{} 0.00015 & 0.12493 \textpm{} 0.00024 & 0.11691 \textpm{} 0.00023 \\
70.0 mm & 0.11665 \textpm{} 0.00016 & 0.13029 \textpm{} 0.00025 & 0.12201 \textpm{} 0.00024 \\
80.0 mm & 0.11883 \textpm{} 0.00016 & 0.13303 \textpm{} 0.00026 & 0.12465 \textpm{} 0.00024 \\
90.0 mm & 0.11988 \textpm{} 0.00016 & 0.13433 \textpm{} 0.00026 & 0.12589 \textpm{} 0.00024 \\
100.0 mm & 0.12033 \textpm{} 0.00016 & 0.13489 \textpm{} 0.00026 & 0.12644 \textpm{} 0.00025 \\
110.0 mm & 0.12051 \textpm{} 0.00016 & 0.13517 \textpm{} 0.00026 & 0.12669 \textpm{} 0.00025 \\
120.0 mm & 0.12055 \textpm{} 0.00016 & 0.13526 \textpm{} 0.00026 & 0.12677 \textpm{} 0.00025 \\ [0.5ex]
\hline \hline 
\textbf{Threshold} & \textbf{1408.0 keV} & \textbf{1836.1 keV} & \\ [0.5ex]
\hline
10.0 mm & 0.00253 \textpm{} 0.00002 & 0.00487 \textpm{} 0.00002 & \\
20.0 mm & 0.01018 \textpm{} 0.00005 & 0.02051 \textpm{} 0.00005 & \\
30.0 mm & 0.02118 \textpm{} 0.00009 & 0.04588 \textpm{} 0.00008 & \\
40.0 mm & 0.02848 \textpm{} 0.00010 & 0.06454 \textpm{} 0.00011 & \\
50.0 mm & 0.03273 \textpm{} 0.00011 & 0.07554 \textpm{} 0.00012 & \\
60.0 mm & 0.03525 \textpm{} 0.00012 & 0.08220 \textpm{} 0.00013 & \\
70.0 mm & 0.03668 \textpm{} 0.00012 & 0.08595 \textpm{} 0.00013 & \\
80.0 mm & 0.03746 \textpm{} 0.00013 & 0.08789 \textpm{} 0.00013 & \\
90.0 mm & 0.03788 \textpm{} 0.00013 & 0.08885 \textpm{} 0.00013 & \\
100.0 mm & 0.03810 \textpm{} 0.00013 & 0.08928 \textpm{} 0.00013 & \\
110.0 mm & 0.03823 \textpm{} 0.00013 & 0.08948 \textpm{} 0.00013 & \\
120.0 mm & 0.03833 \textpm{} 0.00013 & 0.08955 \textpm{} 0.00014 & \\
\hline
\end{tabular}
\caption{P/T ratios vs. spatial grouping threshold}
Temporal threshold set at 250 ns
\label{table:pt_ratios_vs_spatial_grouping_threshold}
\end{table}

\begin{table}[ht]
\centering 
\begin{tabular}{c c c c}
\hline \hline
\textbf{Threshold} & \textbf{121.8.1 keV} & \textbf{344.3 keV} & \textbf{511.0 keV} \\ [0.5ex]
\hline
50 ns & 0.10762 \textpm{} 0.00024 & 0.07470 \textpm{} 0.00019 & 0.13428 \textpm{} 0.00016 \\
100 ns & 0.11031 \textpm{} 0.00024 & 0.07820 \textpm{} 0.00019 & 0.14269 \textpm{} 0.00017 \\
150 ns & 0.11160 \textpm{} 0.00025 & 0.07994 \textpm{} 0.00020 & 0.14770 \textpm{} 0.00017 \\
200 ns & 0.11290 \textpm{} 0.00025 & 0.08178 \textpm{} 0.00020 & 0.15423 \textpm{} 0.00018 \\
250 ns & 0.11332 \textpm{} 0.00025 & 0.08255 \textpm{} 0.00020 & 0.15694 \textpm{} 0.00018 \\
300 ns & 0.11344 \textpm{} 0.00025 & 0.08324 \textpm{} 0.00020 & 0.15897 \textpm{} 0.00018 \\
350 ns & 0.11341 \textpm{} 0.00025 & 0.08350 \textpm{} 0.00020 & 0.15958 \textpm{} 0.00018 \\
300 ns & 0.11341 \textpm{} 0.00025 & 0.08373 \textpm{} 0.00020 & 0.16004 \textpm{} 0.00019 \\
450 ns & 0.11343 \textpm{} 0.00025 & 0.08383 \textpm{} 0.00020 & 0.16023 \textpm{} 0.00019 \\
500 ns & 0.11347 \textpm{} 0.00025 & 0.08393 \textpm{} 0.00020 & 0.16042 \textpm{} 0.00019 \\ [0.5ex]
\hline \hline
\textbf{Threshold} & \textbf{898.0 keV} & \textbf{1173.3 keV} & \textbf{1332.5 keV} \\ [0.5ex]
\hline
50 ns & 0.09379 \textpm{} 0.00013 & 0.10276 \textpm{} 0.00021 & 0.09577 \textpm{} 0.00020 \\
100 ns & 0.10356 \textpm{} 0.00014 & 0.11471 \textpm{} 0.00023 & 0.10715 \textpm{} 0.00022 \\
150 ns & 0.10881 \textpm{} 0.00015 & 0.12089 \textpm{} 0.00024 & 0.11298 \textpm{} 0.00022 \\
200 ns & 0.11443 \textpm{} 0.00015 & 0.12752 \textpm{} 0.00025 & 0.11932 \textpm{} 0.00023 \\
250 ns & 0.11665 \textpm{} 0.00016 & 0.13029 \textpm{} 0.00025 & 0.12201 \textpm{} 0.00024 \\
300 ns & 0.11825 \textpm{} 0.00016 & 0.13230 \textpm{} 0.00025 & 0.12401 \textpm{} 0.00024 \\
350 ns & 0.11869 \textpm{} 0.00016 & 0.13282 \textpm{} 0.00026 & 0.12452 \textpm{} 0.00024 \\
400 ns & 0.11901 \textpm{} 0.00016 & 0.13321 \textpm{} 0.00026 & 0.12488 \textpm{} 0.00024 \\
450 ns & 0.11914 \textpm{} 0.00016 & 0.13336 \textpm{} 0.00026 & 0.12503 \textpm{} 0.00024 \\
500 ns & 0.11925 \textpm{} 0.00016 & 0.13349 \textpm{} 0.00026 & 0.12514 \textpm{} 0.00024 \\ [0.5ex]
\hline \hline 
\textbf{Threshold} & \textbf{1408.0 keV} & \textbf{1836.1 keV} & \\ [0.5ex]
\hline
50 ns & 0.02944 \textpm{} 0.00011 & 0.06653 \textpm{} 0.00011 \\
100 ns & 0.03269 \textpm{} 0.00012 & 0.07500 \textpm{} 0.00012 \\
150 ns & 0.03427 \textpm{} 0.00012 & 0.07915 \textpm{} 0.00012 \\
200 ns & 0.03598 \textpm{} 0.00012 & 0.08386 \textpm{} 0.00013 \\
250 ns & 0.03668 \textpm{} 0.00012 & 0.08595 \textpm{} 0.00013 \\
300 ns & 0.03714 \textpm{} 0.00013 & 0.08757 \textpm{} 0.00013 \\
350 ns & 0.03724 \textpm{} 0.00013 & 0.08800 \textpm{} 0.00013 \\
400 ns & 0.03732 \textpm{} 0.00013 & 0.08830 \textpm{} 0.00013 \\
450 ns & 0.03736 \textpm{} 0.00013 & 0.08841 \textpm{} 0.00013 \\
500 ns & 0.03739 \textpm{} 0.00013 & 0.08850 \textpm{} 0.00013 \\
\end{tabular}
\caption{P/T ratios vs. temporal grouping threshold}
Spatial threshold set at 70 mm
\label{table:pt_ratios_vs_temporal_grouping_threshold}
\end{table}

\begin{table}[ht]
\centering 
\begin{tabular}{c c c c c}
& \textbf{Error in} & \textbf{Error in} & \textbf{Error in} & \textbf{Correctly} \\
& \textbf{Hit 1} & \textbf{Hits 1-2} & \textbf{Any Hit} & \textbf{Sequenced} \\ [0.5ex]
\hline
Full-Energy & 33.54 \textpm{} 0.36 \% & 45.98 \textpm{} 0.44 \% & 52.76 \textpm{} 0.48 \% & 47.24 \textpm{} 0.45 \% \\
Partial-Energy & 47.31 \textpm{} 0.44 \% & 58.81 \textpm{} 0.50 \% & 61.13 \textpm{} 0.52 \% & 38.87 \textpm{} 0.38 \% \\
\hline
\end{tabular}
\caption{Probability of sequencing errors for full-energy vs. partial-energy tracks}
1000.0 keV track energy, (3.0 mm, 2.0 keV) detector resolution
\label{table:sequencing_quality_vs_photon_energy}
\end{table}

\begin{table}[ht]
\centering 
\begin{tabular}{c c c c c}
\textbf{Position} & \textbf{Error in} & \textbf{Error in} & \textbf{Error in} & \textbf{Correctly} \\
\textbf{Resolution} & \textbf{Hit 1} & \textbf{Hits 1-2} & \textbf{Any Hit} & \textbf{Sequenced} \\ [0.5ex]
\hline
0.0 mm & 10.15 \textpm{} 0.18 \% & 13.07 \textpm{} 0.21 \% & 14.68 \textpm{} 0.22 \% & 85.32 \textpm{} 0.68 \% \\
0.25 mm & 13.64 \textpm{} 0.21 \% & 18.00 \textpm{} 0.25 \% & 20.78 \textpm{} 0.27 \% & 79.22 \textpm{} 0.64 \% \\
0.50 mm & 17.28 \textpm{} 0.24 \% & 23.06 \textpm{} 0.29 \% & 26.86 \textpm{} 0.31 \% & 73.14 \textpm{} 0.61 \% \\
0.75 mm & 20.32 \textpm{} 0.27 \% & 27.31 \textpm{} 0.32 \% & 31.83 \textpm{} 0.35 \% & 68.17 \textpm{} 0.58 \% \\
1.0 mm & 22.89 \textpm{} 0.29 \% & 30.87 \textpm{} 0.34 \% & 35.96 \textpm{} 0.38 \% & 64.04 \textpm{} 0.55 \% \\
1.5 mm & 26.89 \textpm{} 0.31 \% & 36.70 \textpm{} 0.38 \% & 42.37 \textpm{} 0.42 \% & 57.63 \textpm{} 0.51 \% \\
2.0 mm & 29.75 \textpm{} 0.33 \% & 40.66 \textpm{} 0.41 \% & 46.83 \textpm{} 0.45 \% & 53.17 \textpm{} 0.49 \% \\
2.5 mm & 31.78 \textpm{} 0.35 \% & 43.60 \textpm{} 0.43 \% & 50.07 \textpm{} 0.47 \% & 49.93 \textpm{} 0.47 \% \\
3.0 mm & 33.54 \textpm{} 0.36 \% & 45.98 \textpm{} 0.44 \% & 52.76 \textpm{} 0.48 \% & 47.24 \textpm{} 0.45 \% \\
4.0 mm & 36.35 \textpm{} 0.38 \% & 49.85 \textpm{} 0.47 \% & 56.70 \textpm{} 0.51 \% & 43.30 \textpm{} 0.42 \% \\
5.0 mm & 38.18 \textpm{} 0.39 \% & 52.53 \textpm{} 0.48 \% & 59.32 \textpm{} 0.52 \% & 40.68 \textpm{} 0.41 \% \\
7.5 mm & 41.05 \textpm{} 0.41 \% & 56.51 \textpm{} 0.51 \% & 63.21 \textpm{} 0.55 \% & 36.79 \textpm{} 0.38 \% \\
10.0 mm & 42.63 \textpm{} 0.42 \% & 58.98 \textpm{} 0.52 \% & 65.55 \textpm{} 0.56 \% & 34.45 \textpm{} 0.37 \% \\
\hline
\end{tabular}
\caption{Probability of sequencing errors vs. position resolution}
1000.0 keV track energy, 2.0 keV energy resolution
\label{table:sequencing_quality_vs_position_resolution}
\end{table}

\begin{table}[ht]
\centering 
\begin{tabular}{c c c c c}
\textbf{Energy} & \textbf{Error in} & \textbf{Error in} & \textbf{Error in} & \textbf{Correctly} \\
\textbf{Resolution} & \textbf{Hit 1} & \textbf{Hits 1-2} & \textbf{Any Hit} & \textbf{Sequenced} \\ [0.5ex]
\hline
0.0 keV & 33.73 \textpm{} 0.36 \% & 45.93 \textpm{} 0.44 \% & 52.59 \textpm{} 0.48 \% & 47.41 \textpm{} 0.45 \% \\
1.0 keV & 33.60 \textpm{} 0.36 \% & 45.95 \textpm{} 0.44 \% & 52.72 \textpm{} 0.48 \% & 47.28 \textpm{} 0.45 \% \\
2.5 keV & 33.59 \textpm{} 0.36 \% & 46.04 \textpm{} 0.44 \% & 52.81 \textpm{} 0.48 \% & 47.18 \textpm{} 0.45 \% \\
5.0 keV & 33.67 \textpm{} 0.36 \% & 46.28 \textpm{} 0.44 \% & 53.11 \textpm{} 0.49 \% & 46.90 \textpm{} 0.45 \% \\
7.5 keV & 34.07 \textpm{} 0.36 \% & 46.72 \textpm{} 0.45 \% & 53.67 \textpm{} 0.49 \% & 46.33 \textpm{} 0.44 \% \\
10.0 keV & 34.60 \textpm{} 0.37 \% & 47.44 \textpm{} 0.45 \% & 54.47 \textpm{} 0.49 \% & 45.53 \textpm{} 0.44 \% \\
\hline
\end{tabular}
\caption{Probability of sequencing errors vs. energy resolution}
1000.0 keV track energy, 3.0 mm position resolution
\label{table:sequencing_quality_vs_energy_resolution}
\end{table}

\begin{table}[ht]
\centering 
\begin{tabular}{c c c c c}
\textbf{Track} & \textbf{Error in} & \textbf{Error in} & \textbf{Error in} & \textbf{Correctly} \\
\textbf{Length} & \textbf{Hit 1} & \textbf{Hits 1-2} & \textbf{Any Hit} & \textbf{Sequenced} \\ [0.5ex]
\hline
2 hits & 15.31 \textpm{} 0.62 \% & 15.31 \textpm{} 0.62 \% & 15.31 \textpm{} 0.62 \% & 84.69 \textpm{} 1.85 \% \\
3 hits & 28.18 \textpm{} 0.66 \% & 38.88 \textpm{} 0.80 \% & 38.88 \textpm{} 0.80 \% & 61.12 \textpm{} 1.08 \% \\
4 hits & 38.03 \textpm{} 0.74 \% & 52.56 \textpm{} 0.91 \% & 56.20 \textpm{} 0.96 \% & 43.80 \textpm{} 0.81 \% \\
5 hits & 40.67 \textpm{} 0.87 \% & 56.89 \textpm{} 1.09 \% & 70.20 \textpm{} 1.26 \% & 29.80 \textpm{} 0.72 \% \\
6 hits & 40.68 \textpm{} 1.14 \% & 58.34 \textpm{} 1.45 \% & 80.79 \textpm{} 1.82 \% & 19.21 \textpm{} 0.72 \% \\
\hline
(All) & 33.54 \textpm{} 0.36 \% & 45.98 \textpm{} 0.44 \% & 52.76 \textpm{} 0.48 \% & 47.24 \textpm{} 0.45 \% \\
\end{tabular}
\caption{Probability of sequencing errors vs. track length}
1000.0 keV track energy, (3.0 mm, 2.0 keV) detector resolution
\label{table:sequencing_quality_vs_track_length}
\end{table}

\begin{table}[ht]
\centering 
\begin{tabular}{c c c c c}
\textbf{FoM} & \textbf{Error in} & \textbf{Error in} & \textbf{Error in} & \textbf{Correctly} \\
\textbf{Type} & \textbf{Hit 1} & \textbf{Hits 1-2} & \textbf{Any Hit} & \textbf{Sequenced} \\ [0.5ex]
\hline 
Type 1 & 40.22 \textpm{} 0.43 \% & 47.91 \textpm{} 0.49 \% & 52.25 \textpm{} 0.52 \% & 47.75 \textpm{} 0.49 \% \\
Type 2 & 33.28 \textpm{} 0.38 \% & 45.11 \textpm{} 0.47 \% & 51.27 \textpm{} 0.51 \% & 48.73 \textpm{} 0.49 \% \\
Type 3 & 56.16 \textpm{} 0.54 \% & 62.47 \textpm{} 0.58 \% & 67.91 \textpm{} 0.62 \% & 32.09 \textpm{} 0.38 \% \\
Type 4 & 13.79 \textpm{} 0.23 \% & 34.10 \textpm{} 0.39 \% & 43.44 \textpm{} 0.46 \% & 56.56 \textpm{} 0.54 \% \\
\end{tabular}
\caption{Performance of multiple FoM definitions}
\label{table:sequencing_quality_vs_FoM_definition}
\end{table}

\end{document}